\documentclass[a4paper]{article}
\usepackage{ISCSLP2026}
\usepackage{ifthen}
\usepackage{graphicx}
\usepackage{booktabs}
\usepackage{array}
\usepackage{amsmath}
\usepackage{tikz}
\usetikzlibrary{arrows.meta,positioning,calc,fit,backgrounds}

\makeatletter
\def\thebibliography#1{\section{References}\eightpt\fontsize{8}{8.6}\selectfont\list
 {[\arabic{enumi}]}{\settowidth\labelwidth{[#1]}\leftmargin\labelwidth
 \advance\leftmargin\labelsep
 \itemsep=0pt \parsep=0pt \topsep=2pt \partopsep=0pt
 \usecounter{enumi}}
 \def\newblock{\hskip .11em plus .33em minus .07em}
 \sloppy\clubpenalty4000\widowpenalty4000
 \sfcode`\.=1000\relax}
\makeatother
\newboolean{blind}
\setboolean{blind}{false}

\title{GenTraceBench: A Benchmark for Tracing Audio Deepfakes \\ Across Pre- and Post-training Stages}
\name{
	\ifthenelse{\boolean{blind}}{Anonymous to ISCSLP}
	{Li Wang$^1$, Kunyu Feng$^1$, Wan Lin$^1$, Dekun Chen$^1$, Qinke Ni$^1$,\\
	 Xueyao Zhang$^1$, Lei Wang$^3$, Jie Shi$^3$, Haizhou Li$^1$, Zhizheng Wu$^{1,2}$}
}
\address{
  \ifthenelse{\boolean{blind}}{Anonymous to ISCSLP}
  {
  	$^1$The Chinese University of Hong Kong, Shenzhen, China\\
    $^2$Amphion Technology Co., Ltd., China\\
    $^3$Huawei Technologies Co., Ltd., China
  }
}
\email{
	\ifthenelse{\boolean{blind}}{Anonymous to ISCSLP}
	{liwang1@link.cuhk.edu.cn, wuzhizheng@cuhk.edu.cn}
}

\begin{document}

\maketitle

\begin{abstract}
Modern text-to-speech (TTS) systems are rarely deployed as unchanged pre-trained models. They are often adapted through supervised fine-tuning (SFT) or preference optimization such as DPO and GRPO. This raises a practical question for audio deepfake forensics: do fingerprints learned from a foundation generator remain valid after adaptation? We present GenTraceBench, a controlled benchmark spanning five TTS architectures, 16 pre-/post-training variants, and 49,728 utterances generated with fixed texts and speaker prompts. Under a train-on-foundation, test-on-adapted protocol, we evaluate binary detection, closed-set attribution, and open-set verification. DPO and GRPO generally preserve fingerprints, whereas some SFT and pre-training-data changes cause substantial drift; effect sizes vary across three forensic backbones. Repeated training runs confirm the largest W2V-BERT attribution drop, while a data-mixture control with comparable speech quality shows that composition change need not cause drift. In W2V-BERT verification, multi-shot enrollment reduces EER for the SFT condition from 44.4\% to 11.0\%, whereas the SingNet-only condition remains at or above 45\% EER.
\end{abstract}
\noindent\textbf{Index Terms}: audio deepfake detection, source tracing, attribution, deepfake verification, text-to-speech, benchmark

\section{Introduction}

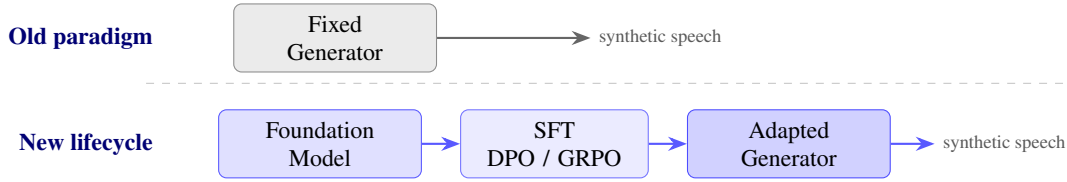
\begin{figure*}[t]
\centering
\begin{tikzpicture}[
  font=\small,
  >={Stealth[length=2.5mm]},
  every node/.style={align=center},
  box/.style={rounded corners=3pt, draw, inner sep=4pt, minimum height=9mm, text=black},
  mono/.style={box, draw=black!45, fill=black!8, text width=2.4cm},
  fnd/.style={box, draw=blue!65, fill=blue!12, text width=2.4cm},
  post/.style={box, draw=blue!55, fill=blue!8, text width=2.2cm},
  adapt/.style={box, draw=blue!70, fill=blue!18, text width=2.4cm},
  lbl/.style={font=\small\bfseries, text=blue!45!black},
  wave/.style={font=\scriptsize, text=black!65}
]
\node[lbl,anchor=east] at (-6.0,0.65) {Old paradigm};
\node[mono] (mono) at (-3.7,0.65) {Fixed\\Generator};
\node[wave] (ow) at (0.6,0.65) {synthetic speech};
\draw[->,thick,black!60] (mono) -- (ow);
\draw[gray!45,dashed] (-6.2,0.05) -- (6.2,0.05);
\node[lbl,anchor=east] at (-6.0,-0.75) {New lifecycle};
\node[fnd] (fnd) at (-3.9,-0.75) {Foundation\\Model};
\node[post] (post) at (-0.8,-0.75) {SFT\\DPO / GRPO};
\node[adapt] (adapted) at (2.3,-0.75) {Adapted\\Generator};
\node[wave] (nw) at (5.15,-0.75) {synthetic speech};
\draw[->,thick,blue!65] (fnd) -- (post);
\draw[->,thick,blue!65] (post) -- (adapted);
\draw[->,thick,blue!65] (adapted) -- (nw);
\end{tikzpicture}
\caption{TTS has shifted from fixed generators to a foundation-then-adapt lifecycle. GenTraceBench studies whether forensic fingerprints remain stable as a foundation model is adapted by SFT or preference-based RL before deployment.}
\label{fig:paradigm}
\end{figure*}

Modern TTS systems are commonly developed in two stages~\cite{zhang2023speechgpt,zhang2024speechalign,zhang-etal-2025-advancing-zero}. A model is first pre-trained for general acoustic modeling. It is then post-trained to improve controllability, quality, or instruction following. Pre-trained models~\cite{wang2023neural,anastassiou2024seedttsfamilyhighqualityversatile,du2024cosyvoice} provide strong starting points. Post-training methods such as supervised fine-tuning (SFT)~\cite{yang2024instructtts} and preference-based alignment~\cite{rafailov2024directpreferenceoptimizationlanguage}, including DPO and GRPO, further refine prosody, speaking style, and instruction adherence. As a result, deployed systems are often not the released pre-trained checkpoints, but adapted descendants with different acoustic behavior (Fig.~\ref{fig:paradigm}).

Evaluation benchmarks for audio deepfake forensics have not kept pace with this lifecycle. Most existing corpora study fixed, fully trained generators~\cite{258dce4d760e4b0885bfe554a1b8d725,yamagishi2021asvspoof,wang2024asvspoof,frank2023wavefake,muller2024mlaad}; attribution and open-set tracing likewise compare finished systems~\cite{dipierno2025reliableaudiodeepfakeattribution,klein2025opensetsourcetracingaudio,wang2025audiodeepfakeverification}. Such static evaluations cannot determine whether model-specific artifacts captured by forensic front ends survive the generator's own adaptation. Here, \emph{fingerprint drift} denotes systematic changes in those artifacts. Closer controlled-shift benchmarks include ShiftySpeech, which varies speaker, language, acoustics, and synthesis distribution~\cite{garg2025shiftyspeech}, and STOPA, which systematically varies acoustic models, vocoders, parameters, and pretrained weights~\cite{firc2025stopa}. These resources target cross-system or component variation. GenTraceBench complements them by following matched checkpoints within a generator lineage while holding prompts and speakers fixed, thereby isolating \emph{generator-training-stage fingerprint drift}.

Related text-generation studies find that RLHF or instruction tuning can alter detector behavior~\cite{xu2025rlhfdetectability,xu2026basemodels}, but do not establish the corresponding effect for speech generators. Ge et al.~\cite{ge2025posttrainingdetector} use ``post-training'' for adapting the \emph{detector}; here the adapted object is the \emph{generator}. These distinctions motivate evaluating forensic robustness across a generator's SFT/RL lifecycle. Our contribution is the controlled benchmark and evaluation protocol, rather than a new detector.

\begin{figure*}[t]
\centering
\begin{tikzpicture}[
  font=\small,
  >={Stealth[length=2.5mm]},
  every node/.style={align=center},
  hdr/.style={font=\bfseries\large, text=blue!45!black},
  ibox/.style={rounded corners=3pt, draw=blue!55, line width=0.7pt, inner sep=4pt, text width=4.0cm, minimum height=11mm},
  proc/.style={rounded corners=10pt, draw=blue!55, line width=0.9pt},
  obox/.style={rounded corners=10pt, draw=blue!55, line width=0.9pt, fill=blue!6, text width=2.8cm, minimum height=4.3cm},
  ph/.style={font=\bfseries, rounded corners=2pt, draw=blue!40, fill=blue!12, inner sep=4pt},
  brl/.style={font=\scriptsize, text=black}
]
\node[hdr] at (2.3,4.7) {Input};
\node[hdr] at (7.8,4.7) {Process};
\node[hdr] at (15.3,4.7) {Output};
\node[ibox] (m)  at (2.3,3.5) {\textbf{Model}\\[1pt]\footnotesize CosyVoice2~\cite{du2024cosyvoice}, F5-TTS~\cite{Chen2025}, FlexiVoice~\cite{chen2026flexivoice}, MaskGCT~\cite{Wang2025}, Vevo2~\cite{zhang2025vevo2unifiedcontrollableframework}};
\node[ibox] (pd) at (2.3,2.0) {\textbf{Pre-training Data}\\[1pt]\footnotesize Emilia~\cite{he2024emiliaextensivemultilingualdiverse}, SingNet~\cite{gu2025singnetlargescalediverseinthewild}};
\node[ibox] (od) at (2.3,0.7) {\textbf{Post-training Data}\\[1pt]\footnotesize INTP~\cite{zhang-etal-2025-advancing-zero}, NVSpeech~\cite{liao2025nvspeechintegratedscalablepipeline}};
\node[ibox] (pr) at (2.3,-0.5) {\textbf{Prompts}\\[1pt]\footnotesize Seed-TTS Eval~\cite{anastassiou2024seedttsfamilyhighqualityversatile}};
\coordinate (bus) at (4.7,1.5);
\foreach \n in {m,pd,od,pr} \draw[blue!55] (\n.east) -| (bus);
\draw[->,blue!55,thick] (bus) -- (5.2,1.5);
\node[proc, minimum width=5.2cm, minimum height=5.0cm] (pc) at (7.8,1.5) {};
\node[ph] (p1) at (6.6,1.5) {Phase 1:\\Pre-training};
\node[ph] (p2) at (9.0,1.5) {Phase 2:\\Post-training};
\draw[->,thick] (p1) -- (p2);
\draw[blue!55,thick] (p2.east) -- (10.8,1.5);
\foreach \y/\txt in {3.3/{General Alignment: DPO}, 2.0/{Modality Expansion: SFT}, 0.7/{Multi-stage: DPO$\to$GRPO}} {
  \draw[blue!55,thick] (10.8,1.5) -- (10.8,\y);
  \draw[->,thick,blue!55] (10.8,\y) -- node[brl,above,pos=0.5]{\txt} (13.9,\y);
}
\draw[->,thick,blue!55] (10.4,-0.6) -- node[brl,above,pos=0.5]{Pre-trained only} (13.9,-0.6);
\draw[blue!55,thick] (p1.south) -- (6.6,-0.6) -- (10.4,-0.6);
\node[obox] (o) at (15.3,1.35) {\textbf{\large GenTrace\\Benchmark}\\[3pt]\footnotesize 16 variants\\49{,}728 utterances};
\end{tikzpicture}
\caption{GenTraceBench construction. Five TTS architectures are combined with controlled pre-/post-training data and fixed Seed-TTS prompts. Each model is traced from its pre-trained foundation (Phase~1) to representative post-training paths (Phase~2), including general alignment (DPO), modality expansion (SFT), and multi-stage evolution (DPO$\to$GRPO). Together with the pre-trained-only path, this yields 16 variants generated from identical texts and speakers.}
\label{fig:construction}
\end{figure*}
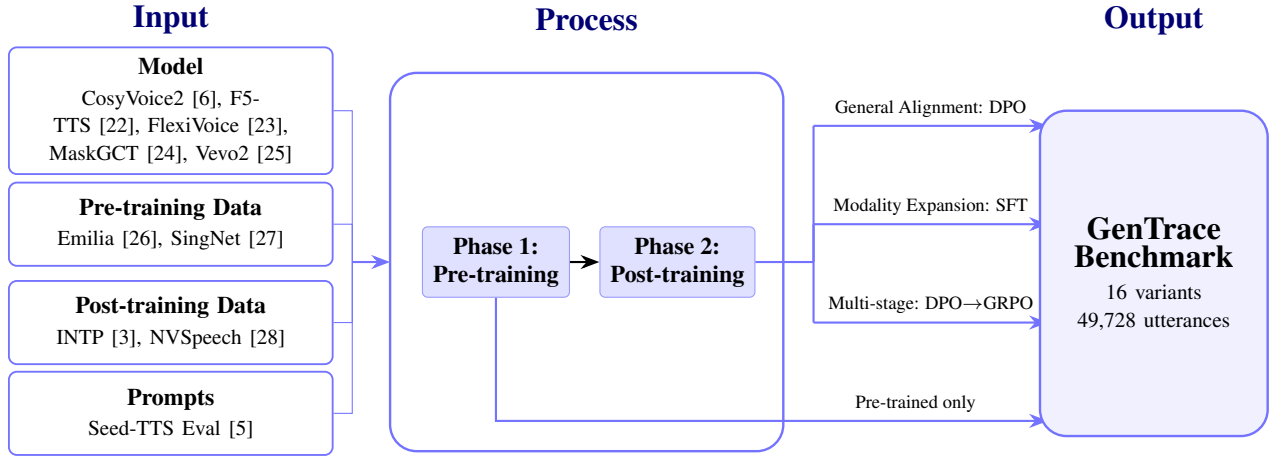

To address this gap, GenTraceBench tracks TTS evolution from pre-training through SFT, RL alignment, multi-stage pipelines, and pre-training-data composition across five architectures. All 16 variants synthesize identical texts with identical speaker prompts, reducing variation due to linguistic content and speaker identity when comparing generators and training stages. Our contributions are:
\begin{itemize}
\item To our knowledge, we construct the first audio benchmark focused on matched, within-lineage generator checkpoints across training stages (16 variants; 49,728 utterances), complementing STOPA's cross-component coverage.
\item We evaluate three complementary forensic tasks under a realistic train-on-foundation, test-on-adapted protocol: binary detection, closed-set attribution, and open-set deepfake verification.
\item In W2V-BERT space, we separate variance-driven CosyVoice2 SFT drift, reduced by prototype averaging, from structural SingNet-only drift.
\item Repeated W2V-BERT training and comparisons with WavLM and AASIST assess robustness across initializations and backbones; an Emilia+SingNet control with comparable speech quality further examines the effect of data composition.
\end{itemize}

\section{GenTraceBench Construction}

\subsection{Generators and Adaptation Framework}

GenTraceBench comprises five representative speech generation architectures evaluated across both pre-training and post-training stages: CosyVoice2~\cite{du2024cosyvoice}, F5-TTS~\cite{Chen2025}, FlexiVoice~\cite{chen2026flexivoice}, MaskGCT~\cite{Wang2025}, and Vevo2~\cite{zhang2025vevo2unifiedcontrollableframework}. These models cover several acoustic modeling paradigms, including LLM-conditioned flow matching, diffusion-based flow matching, autoregressive token modeling with flow-matching vocoders, and masked generative modeling.

\noindent\textbf{Pre-training.}
Most models are pre-trained on the large-scale Emilia corpus~\cite{he2024emiliaextensivemultilingualdiverse}, except CosyVoice2, which uses its official training corpus. To probe the effect of pre-training data composition, we additionally include Vevo2 variants pre-trained on the singing-voice corpus SingNet~\cite{gu2025singnetlargescalediverseinthewild} alone and on Emilia+SingNet.

\noindent\textbf{Post-training adaptation.}
We consider three settings. \textbf{General alignment} applies DPO using INTP preference data~\cite{zhang-etal-2025-advancing-zero}. \textbf{Modality expansion} fine-tunes CosyVoice2 with SFT on NVSpeech~\cite{liao2025nvspeechintegratedscalablepipeline} to add vocalizations such as laughter. \textbf{Multi-stage evolution} uses DPO$\rightarrow$GRPO for Vevo2 and S1 (multi-modal DPO)$\rightarrow$S2 (decoupling GRPO)$\rightarrow$S3 (instruction GRPO) for FlexiVoice. The DPO scale is architecture-specific: INTP uses $\beta{=}0.1/10/1{,}000$ for CosyVoice2/MaskGCT/F5-TTS, respectively; thus the largest value applies only to F5-TTS. Other adaptations follow their source recipes~\cite{liao2025nvspeechintegratedscalablepipeline,zhang2025vevo2unifiedcontrollableframework,chen2026flexivoice} and are detailed in Table~\ref{tab:setup}.

\subsection{Controlled Generation Protocol}

All variants generate speech from the widely adopted Seed-TTS evaluation corpus~\cite{anastassiou2024seedttsfamilyhighqualityversatile}, which provides diverse speakers and 3,108 utterances (1,088 English and 2,020 Chinese). Identical speaker prompts and target texts are used across all 16 variants, ensuring that any difference captured by a forensic system reflects the generator's architecture, pre-training data, or post-training method, rather than content or speaker variation.

\subsection{Partitioning and Statistics}

The source corpus is split into training (1,035), validation (517), and test (1,556) utterance sets while preserving the roughly 2:1 Chinese-to-English ratio. The same split is applied to every variant so that train/test partitions remain parallel across models. In total, GenTraceBench contains 16 model variants and 49,728 synthesized utterances, together with the corresponding \textit{bona fide} human speech.

\section{Experimental Setup}

\subsection{Task Formulation}

\noindent\textbf{Binary detection} distinguishes bona fide human speech from synthetic speech, formulated as binary classification~\cite{258dce4d760e4b0885bfe554a1b8d725}.

\noindent\textbf{Closed-set attribution} is multi-class classification over a fixed set of known generator families. Given a synthetic sample, the system predicts which family produced it. All foundation families are represented during training; adapted variants are evaluated under their corresponding family labels.

\noindent\textbf{Deepfake verification} determines whether two audio samples originate from the same generator~\cite{wang2025audiodeepfakeverification}. In our open-set evaluation, the verifier is trained on foundation generators and tested on previously unseen adapted variants, measuring fingerprint stability across training stages~\cite{klein2025opensetsourcetracingaudio}.

\subsection{Forensic Systems}

In the main experiments, all three task-specific systems use a pre-trained Wav2Vec2-BERT~\cite{chung2021w2vbertcombiningcontrastivelearning} backbone (580M parameters) that is fine-tuned end-to-end with the task heads.

\noindent\textbf{Detection and attribution.}
For detection, a linear classifier on top of the pooled Wav2Vec2-BERT representation separates bona fide from synthetic speech. For attribution, a linear classifier predicts the source family among the five foundation generators.

\noindent\textbf{Verification.}
The verification system maps each utterance to a 512-dimensional embedding trained with the AAM-Softmax loss (margin 0.3, scale 15) over the five foundation-generator classes. At test time, verification is performed by cosine scoring between embeddings. In the \emph{multi-shot} setting, enrollment and test prototypes are each averaged over $N$ utterances of the same source before scoring, which suppresses utterance-level variance while preserving systematic (mean) shifts of the fingerprint.

\subsection{Training Protocol}
\label{sec:training}

We adopt a \textbf{train-on-foundation, test-on-adapted} protocol. Forensic systems are trained on foundation-generator samples and, for detection, bona fide speech. We evaluate the test split of all variants, including adaptations withheld from training. All systems use AdamW for 100 epochs with batch size 64. By default, checkpoints are selected by validation accuracy. Table~\ref{tab:main_results} additionally reports detection and attribution after three training epochs to examine early-training sensitivity, since the converged detector reaches 99.5\% pooled accuracy and 1.0\% EER. Results from these two checkpoint settings are analyzed separately. Repeated W2V-BERT attribution training uses random seeds 1, 42, and 2026; cross-backbone and embedding comparisons use seed 42. Further hyperparameters are listed in Table~\ref{tab:setup}. We will release GenTraceBench to support lifecycle-aware forensic evaluation.

\subsection{Evaluation Metrics}

For detection and attribution, we report per-variant test accuracy; for verification, equal error rate (EER). We also report Seed-TTS word error rate (WER) and prompt-speaker similarity (SIM-o)~\cite{anastassiou2024seedttsfamilyhighqualityversatile}. Variation across three independent W2V-BERT training runs is summarized by the mean and sample standard deviation. For fixed-seed comparisons, 10,000 paired bootstrap resamples of matched prompt identifiers provide 95\% confidence intervals (CIs) for test-sample variability. These intervals measure a different source of uncertainty from variation across training runs.

\section{Results and Analysis}

Table~\ref{tab:main_results} presents speech-quality measures alongside detection and attribution from the early-training evaluation. Subsequent analyses assess robustness using validation-selected checkpoints.

\subsection{Impact of Adaptation Stage on Detection and Attribution}

In Table~\ref{tab:main_results}, RL-based alignment changes attribution accuracy by at most 1.1 percentage points, although detection changes by up to 2.1 points (MaskGCT). Vevo2 attribution decreases from 92.16\% to 91.07\% after DPO, F5-TTS detection increases from 99.81\% to 100\%, and the three-stage FlexiVoice pipeline remains within 0.5 points of its foundation. These results indicate limited attribution changes under RL alignment. Sec.~\ref{sec:mechanism} examines the corresponding changes in embedding geometry.

Changing Vevo2's pre-training corpus from Emilia to SingNet reduces attribution accuracy in Table~\ref{tab:main_results} from 92.16\% to 84.06\%, a decrease of 8.1 percentage points. For CosyVoice2, SFT on NVSpeech lowers detection accuracy to 87.28\%. At validation-selected checkpoints across three independent runs, mean attribution decreases are $0.13{\pm}0.23$ points for the eight RL variants, $0.64{\pm}0.06$ for CosyVoice2 SFT, and $10.15{\pm}0.97$ for SingNet-only Vevo2. Paired bootstrap intervals exclude zero for the latter two conditions in every run, showing consistent decreases across initializations. Adaptation recipes and $\beta$ values vary across architectures, including $\beta{=}1{,}000$ for F5-TTS, so these comparisons do not isolate the effect of $\beta$.

\begin{table}[t]
\caption{Detection, attribution, and speech quality across generator variants. Accuracies use the early-training evaluation described in Sec.~\ref{sec:training}; WER and SIM-o follow Seed-TTS.}
\label{tab:main_results}
\centering
{\setlength{\tabcolsep}{3pt}%
\resizebox{\linewidth}{!}{
\begin{tabular}{lcccc}
\toprule
\textbf{Variant} & \textbf{Det. Acc.~(\%)$\uparrow$} & \textbf{Attr. Acc.~(\%)$\uparrow$} & \textbf{WER~(\%)$\downarrow$} & \textbf{SIM-o$\uparrow$} \\
\midrule
\textbf{Bona fide} & 92.16 & -- & -- & -- \\
\midrule
\multicolumn{5}{c}{\textbf{\textit{Part I: Post-training adaptation}}}\\
\midrule
\textbf{Vevo2} (pre-trained) & 97.69 & 92.16 & 3.18 & 0.734 \\
\quad DPO & 98.59 & 91.07 & 3.27 & 0.733 \\
\quad GRPO & 97.94 & 91.97 & 3.19 & 0.733 \\
\midrule
\textbf{FlexiVoice} (pre-trained) & 97.17 & 78.73 & 2.52 & 0.710 \\
\quad DPO (S1) & 97.69 & 78.41 & 2.53 & 0.709 \\
\quad S1$\rightarrow$S2 & 97.37 & 79.18 & 2.70 & 0.707 \\
\quad S1$\rightarrow$S2$\rightarrow$S3 & 96.72 & 78.66 & 3.08 & 0.707 \\
\midrule
\textbf{CosyVoice2} (pre-trained) & 89.01 & 99.81 & 1.64 & 0.720 \\
\quad DPO & 87.66 & 99.74 & 1.43 & 0.740 \\
\quad SFT (NVSpeech) & 87.28 & 98.07 & 2.15 & 0.706 \\
\midrule
\textbf{MaskGCT} (pre-trained) & 92.48 & 98.97 & 2.43 & 0.753 \\
\quad DPO & 94.60 & 98.46 & 2.34 & 0.751 \\
\midrule
\textbf{F5-TTS} (pre-trained) & 99.81 & 99.81 & 4.63 & 0.731 \\
\quad DPO & 100.00 & 99.81 & 1.66 & 0.694 \\
\midrule
\multicolumn{5}{c}{\textbf{\textit{Part II: Pre-training data composition (Vevo2)}}}\\
\midrule
Emilia (default) & 97.69 & 92.16 & 3.18 & 0.734 \\
Emilia + SingNet & 97.62 & 92.67 & 5.07 & 0.729 \\
SingNet only & 94.54 & 84.06 & 29.43 & 0.675 \\
\bottomrule
\end{tabular}
}}
\end{table}

\subsection{Robustness across Backbones and Initializations}
\label{sec:crossbackbone}

To assess dependence on the forensic model and its initialization, we repeat W2V-BERT training three times and evaluate WavLM~\cite{chen2022wavlm} and AASIST~\cite{jung2022aasist} under the same data and checkpoint-selection protocol. Every variant is evaluated against its foundation identity using the complete test split.

Table~\ref{tab:crossbackbone} compares attribution changes across backbones. The mean decrease under RL alignment is at most 1.39 percentage points, while CosyVoice2 SFT yields decreases of 0.64--8.16 points. The Emilia+SingNet control retains comparable speech quality (WER 5.07\%, SIM-o 0.729, versus 3.18\% and 0.734 for Emilia) and yields attribution decreases of $-0.26$, $0.26$, and $0.13$ points. The corresponding paired-bootstrap 95\% CIs are $[-1.86,1.41]$, $[-2.25,2.76]$, and $[-1.03,1.29]$ for W2V-BERT, WavLM, and AASIST, respectively; all include zero. SingNet-only reduces W2V-BERT and AASIST accuracy by 10.03 and 3.08 points, whereas WavLM improves by 0.77 points with a CI spanning zero. Thus, changing data composition need not cause drift, although the SingNet-only effect remains confounded by substantial quality loss and depends on the forensic representation.

\subsection{Quality and Detectability}

Speech fidelity and detectability do not have a simple monotonic relationship. CosyVoice2 achieves high fidelity (WER 1.64\%, SIM-o 0.720) and has the lowest foundation-model detection accuracy (89.01\%). After DPO, WER further improves to 1.43\%, while detection decreases to 87.66\%. For this architecture, quality optimization also appears to reduce detectable artifacts. F5-TTS behaves differently. DPO reduces WER from 4.63\% to 1.66\%, but detection still saturates at 100\%. Perceptual improvement therefore does not necessarily imply better evasion. It may also strengthen regularities that detectors can exploit.

\begin{table}[t]
\caption{Attribution accuracy decrease across forensic backbones at validation-selected checkpoints (percentage points; negative values indicate improvement). RL averages eight variants; SFT denotes CosyVoice2/NVSpeech. Vevo2 data conditions use Emilia as the reference.}
\label{tab:crossbackbone}
\centering
\resizebox{\linewidth}{!}{%
\begin{tabular}{lccc}
\toprule
\textbf{Condition} & \textbf{W2V-BERT} & \textbf{WavLM} & \textbf{AASIST} \\
\midrule
RL alignment (DPO/GRPO) & 0.17 & 1.39 & 1.27 \\
SFT (NVSpeech) & 0.64 & 5.01 & 8.16 \\
Emilia + SingNet & $-0.26$ & 0.26 & 0.13 \\
SingNet only & 10.03 & $-0.77$ & 3.08 \\
\bottomrule
\end{tabular}}
\end{table}

\subsection{Attribution Ambiguity from Shared Lineage}

The attribution results suggest that most generator families retain distinguishable signatures after adaptation. FlexiVoice is an exception, with accuracy around 79\% in Table~\ref{tab:main_results}; 99\% of its attribution errors are assigned to Vevo2. The models share pre-training data (Emilia) and a flow-matching/Vocos vocoder backend. Their embeddings also have the highest cross-family cosine similarity among the evaluated pairs. These observations suggest that shared data and design choices may produce overlapping forensic representations, although the present comparison does not isolate their individual effects.

\subsection{Mechanism: Variance-Driven vs.\ Structural Drift}
\label{sec:mechanism}

Detection and attribution also depend on decision boundaries, so we analyze the attribution embedding space directly. Relative to each foundation, \emph{centroid shift} measures displacement of the class mean, while the \emph{change in within-class dispersion} measures how the spread around that mean varies. Averaging multiple samples can reduce within-class variability but does not remove a systematic shift of the mean. Table~\ref{tab:geometry} reports both measures at validation-selected checkpoints.

The Emilia+SingNet mixture produces only small changes in both measures. SingNet-only increases dispersion for all three backbones and shifts the W2V-BERT and WavLM centroids, indicating a change in the embedding distribution even where WavLM accuracy improves. For W2V-BERT, SFT leaves the centroid nearly unchanged while increasing dispersion, whereas SingNet-only changes both measures. The decomposition depends on the representation: SFT also increases WavLM dispersion, while AASIST exhibits a larger centroid shift and reduced dispersion.

The multi-shot verification results are consistent with this decomposition. In W2V-BERT space, the in-domain EER falls from 5.2\% to 1.6\% and the RL-condition EER from 14.0\% to 10.9\% as $N$ grows from 1 to 10. CosyVoice2 SFT improves from 44.4\% to 11.0\%, consistent with reduced within-class variability. SingNet-only Vevo2 remains at or above 45\% EER for all tested $N$, consistent with a systematic centroid shift. Multi-sample enrollment is therefore effective for this SFT condition with W2V-BERT; evaluating whether this holds more broadly requires additional generators and backbones.

\begin{table}[t]
\caption{Changes in embedding geometry at validation-selected checkpoints. $C$: centroid shift; $S$: change in within-class dispersion (cosine distance, $\times10^{-2}$). Conditions follow Table~\ref{tab:crossbackbone}.}
\label{tab:geometry}
\centering
{\setlength{\tabcolsep}{3pt}\small
\begin{tabular}{@{}lcccccc@{}}
\toprule
& \multicolumn{2}{c}{\textbf{W2V-BERT}} & \multicolumn{2}{c}{\textbf{WavLM}} & \multicolumn{2}{c}{\textbf{AASIST}} \\
\cmidrule(lr){2-3}\cmidrule(lr){4-5}\cmidrule(lr){6-7}
\textbf{Condition} & $C$ & $S$ & $C$ & $S$ & $C$ & $S$ \\
\midrule
RL alignment & 0.01 & $-0.05$ & 0.09 & 0.69 & 0.08 & 0.19 \\
SFT (NVSpeech) & 0.01 & 0.72 & 0.15 & 3.74 & 0.65 & $-1.65$ \\
Emilia + SingNet & 0.00 & $-0.04$ & 0.00 & 0.11 & 0.00 & $-0.06$ \\
SingNet only & 0.55 & 6.52 & 0.62 & 1.52 & 0.08 & 1.98 \\
\bottomrule
\end{tabular}}
\end{table}

\begin{table}[t]
\caption{Training configurations for forensic systems and TTS adaptation. Post-training settings follow~\cite{zhang-etal-2025-advancing-zero,zhang2025vevo2unifiedcontrollableframework,chen2026flexivoice}.}
\label{tab:setup}
\centering
{\setlength{\tabcolsep}{3pt}%
\begin{tabular}{p{0.30\linewidth}>{\raggedright\arraybackslash}p{0.64\linewidth}}
\toprule
\textbf{Component} & \textbf{Setting} \\
\midrule
Forensic training & AdamW; 100 epochs; batch size 64; learning rate $10^{-5}$ (SSL), $10^{-4}$ (AASIST); validation-based selection \\
Verification head & 512-d SSL / 128-d AASIST; AAM-Softmax ($m{=}0.3$, $s{=}15$) \\
INTP DPO & 1 epoch; inverse-square-root schedule; 4,000 warmup steps; $\beta$: CosyVoice2 0.1, MaskGCT 10, F5-TTS 1{,}000 \\
Vevo2 stages & DPO$\rightarrow$GRPO; learning rate $5{\times}10^{-6}$; group size 8; KL coefficient 0.1; 9,000 updates \\
FlexiVoice stages & DPO$\rightarrow$GRPO; learning rate $10^{-5}$; $\beta{=}0.1$; group size $8{\to}6$; 2--3 epochs \\
\bottomrule
\end{tabular}}
\end{table}

\newpage
\section{Conclusions}

We presented GenTraceBench, a controlled benchmark across TTS training stages. Repeated W2V-BERT training and cross-backbone comparisons show small average attribution changes under RL alignment, backbone-dependent SFT effects, and a reproducible SingNet-only shift. The Emilia+SingNet control shows that changing data composition need not cause drift. W2V-BERT geometry and multi-shot verification further distinguish the evaluated SFT and SingNet-only conditions.

These findings suggest that deployment decisions should account for the type of fingerprint change. When adaptation primarily increases within-generator variability, averaging several enrollment samples may improve verification. When it shifts the mean representation, additional samples may be insufficient, motivating updates to the forensic system. Evaluation should therefore track generator lineages across training stages.

The study is limited to one SFT configuration, verification training on five generator families, repeated training only for W2V-BERT, architecture-specific adaptation hyperparameters, and clean Chinese and English audio. Broader coverage of generators, initializations, languages, and recording conditions remains future work. GenTraceBench will be released publicly.

\section{Acknowledgements}

This work is supported by the Program for Guangdong Introducing Innovative and Entrepreneurial Teams under Grant No.~2023ZT10X044. The work of Zhizheng Wu is supported by the National Natural Science Foundation of China under Grant No.~62376237.

\clearpage
\bibliographystyle{IEEEtran}
\bibliography{gentrace}

\end{document}